# Relaxation processes of densified silica glass

Antoine Cornet,[1,a)] Valérie Martinez,[1] Dominique de Ligny,[2] Bernard Champagnon[1] and Christine Martinet[1]

[1]*Univ Lyon, Université Claude Bernard Lyon 1, CNRS, Institut Lumière Matière, F-69622, VILLEURBANNE, France*

[2]*Dept. Werkstoffwissenschaften, Lehrstuhl für Glas und Keramik, Universität Erlangen-Nürnberg, Martensstr. 5, D-91058 Erlangen, Germany*

Densified $SiO_2$ glasses, obtained from different pressure and temperature routes have been annealed over a wide range of temperature far below the glass transition temperature (500°C-900°C). Hot and cold compressions were useful to separate the effects of pressure and the compression temperature. In-situ micro-Raman spectroscopy was used to follow the structural evolution during the thermal relaxation. A similar glass structure between the non-densified silica and the recovered densified silica after the temperature annealing demonstrates a perfect recovery of the non-densified silica glass structure. While the density decreases monotonically, the structural relaxation takes place through a more complex mechanism, which shows that density is not a sufficient parameter to fully characterize the structure of densified silica glass. The relaxation takes place through a transitory state, consisting in an increase of the network inhomogeneity, shown by an increase in intensity of the D2 band which is associated with 3 membered rings. The activation energy of these processes is 255±45 kJ/mol for the hot compressed samples. The kinetic is overall faster for the cold compressed samples. In that last case the relaxation is partially activated by internal stresses release.

[a)] Author to whom correspondence should be addressed. Electronic mail: antoine.cornet@univ-lyon1.fr

## I. INTRODUCTION

Because its importance to several fields as geology, telecommunications, industry, silica glass has been without doubt the most studied material, with a considerable documentation. While the inner structure has been characterized really soon in the early studies for the short range order, the network formed by the $SiO_4$ tetrahedral units reveals a greater complexity. From the initial study of Bridgman[1] in 1953, it is known than silica undergoes a permanent densification after a compression above 10 GPa. Since this initial work, considerable efforts have been made on the densification of silica glass. From High Pressure and High Temperature (HP-HT) runs,[2] it has been established that the application of temperature during the compression leads to higher densification ratios compared to room temperature for a given applied pressure.[3] The different mechanical and structural responses under high pressure have been clarified, introducing two general features: the mid-long range order modifications and an increase of the silicon coordination in the short range order. Hemley et al. demonstrated a narrowing of the inter tetrahedral angle distribution, a decrease of the mean value of this angle and an enhancement of the number of small rings.[4] This was later confirmed by several in situ studies under high pressure[5], including numerical simulations.[6] The silicon coordination number increase has been demonstrated from infrared spectroscopy[7] and XRD[8] measurements.

Long time after the first ex-situ HP-HT runs, a more careful and quantitative approach on such processes was shown. Poe et al. demonstrated that the application of high temperatures during compression affects mainly inter tetrahedral angles, while $SiO_4$ units are quite preserved.[9] In addition, numerical simulations[10] predicted a broadening of the ring size statistic, with an increase in the proportion of the rings composed by more than 6 tetrahedrons. However, the proportion of 3 and 4 membered rings remains unchanged. Trachenko et al. suggested a first window where the topological modifications appear in the pressure range. They showed in a numerical study that rebondings events begin between 3 GPa and 5 GPa at 300K.[11] Inamura extended the previous observation experimentally, in a remarkable work,[12] where the thermodynamic region between 0-20 GPa and 0- 1200°C was scanned in XRD experiments, and the high coupling temperature-pressure range is clearly highlighted. In all cases, the mid-range order modifications, after compression-decompression cycle above the elastic limit, remain irreversible for the recovered sample.

Despite the considerable amount of studies and data on in-situ and ex-situ compression of silica, very low interest was shown for the reverse transformation, i.e. the relaxation during high temperature annealing, from high density to low density, isothermally or not. It is yet quite appropriate for in situ measurements, since the application of high temperatures is generally more easily suitable than the application of high pressures. Using such experimental methods, Hummel and Arndt evidenced a bump in the dependence of the refractive index to the density in silica, demonstrated an anormal state[13] in the linear relation. More recently, Surovtsev et al. investigated both the boson peak and density changes during the temperature annealing from hot compressed silica samples.[14] In our study, in-situ Raman experiments at different fixed temperature have been performed on silica glass from several initial densifications in order to study the relaxation processes. From isothermal treatments, the activation energies of the density relaxation process have been deduced and discussed. Depending on the macroscopic density and initial structure of silica glasses, the structural modifications implied in the transformation are detailed.

It is known that the density evolution during the compression is supported by different kinds of structural changes, reviewed in Ref. 15. Because of this, no direct correlation between a spectroscopic signature and the density can be found in the whole pressure range. Nevertheless, the interest of a spectroscopic signature as a density marker has been a matter of debate for the recovered samples.[16,17] Recently, Martinet et al. suggested a satisfying correlation between the density of the recovered sample and the width of the main band of the corresponding Raman spectrum[18] for several HP-HT densifications. This result connects the recovered density to the distribution of the Si-O-Si angles. In this study we will discuss the consistency of this result during the isothermal relaxation. The relation between the Raman spectra and the density established in Ref. 18 is here further tested.

At the end of the discussion we will try to extend our results to a more generalized vision of glass in terms of polyamorphism and presence of internal stresses.

## II. EXPERIMENTAL SET-UP

### A. Sample preparation : densification process

Using commercial Suprasil 300 ([OH] < 1 ppm) as initial $SiO_2$ glass, six densifications were performed (see Table I) and some of their related Raman spectra shown in Fig. 1. Three samples were synthesized in a high pressure belt-type press apparatus, as reported elsewhere.[18] The samples were compacted in small cylindric pellets (4 mm diameter and 6 mm length) and placed into a closed hexagonal boron nitride crucible. The entire assembly was introduced inside the carbon tube heater inserted in the pyrophyllite HP gasket. The HP-HT conditions of the synthesis are listed in the Table I. The pressure was first increased up to the desired value, and then the sample was slowly heated up (≈1°C/s) and maintained at the desired temperature for 10 minutes, and then cooled to room temperature. Finally, the pressure was released to room pressure and the sample recovered. The macroscopic densities were determined using buoyancy measurement in Toluene.

Three other silica Suprasil 300 samples were densified using a Chervin-type Diamond Anvil Cell (DAC) to provide pure hydrostatic compression as it has been previously detailed.[19] The experimental volume corresponds to a hole in the metal gasket of diameter 200 µm. Suprasil type 300 samples and ruby spheres were loaded. The transmitting medium was 16:3:1 water ethanol methanol mixture, which remains hydrostatic until 10.5 GPa,[20] and quasi-hydrostatic up to 20 GPa.

Pressure was determined in-situ from the emission of $^2E$-$^4A_2$ ($R_1$ line) $Cr^{3+}$ transition of ruby spheres introduced in the experimental volume.[21] Maximal pressure was maintained during twenty minutes. The pressure stabilization was checked by measuring the ruby luminescence before and after the experiment. No significant modification was observed. The densification rate is deduced from the density-maximal pressure reached calibration curve.[22]

TABLE I. Conditions of P,T process and densities obtained for recovered samples. The density for Belt press samples were obtained from buoyancy measurement in Toluene at room temperature and for DAC samples from the density/Pmax calibration curve.[22]

| | **Pressure** | **Temperature** | **Density (g.cm$^{-3}$)** | **Densification ratio** | **Sample name** |
|---|---|---|---|---|---|
| **Belt apparatus** | 5 GPa | 425 °C | d = 2.40 ± 0.03 | Δρ/ρ = 9.2% | Belt 9.2% |
| | 5 GPa | 750 °C | d = 2.54 ± 0.03 | Δρ/ρ = 15.5% | Belt 15.5% |
| | 5 GPa | 1020 °C | d = 2.56 ± 0.02 | Δρ/ρ = 16.5% | Belt 16.5% |
| **Diamond anvil cell** | 13,8 GPa | 25 °C | d = 2.46 ± 0.04 | Δρ/ρ = 12% | DAC 12% |
| | 16 GPa | 25 °C | d = 2.55 ± 0.04 | Δρ/ρ = 16% | DAC 16% |
| | 25 GPa | 25 °C | d = 2.66 ± 0.02 | Δρ/ρ = 21% | DAC 21% |

## B. Heating and spectra collection

The Raman spectra were recorded with a Renishaw RM 1000 spectrometer with a resolution of 1.6 $cm^{-1}$, using the 532 nm excitation of a $Nd^{3+}$:YAG laser. A piece of glass was introduced in a Linkam TMS 1500 device. The Linkam cell was set under a ULWD x 20 Olympus objective with an incident laser power below 15 mW on the sample. The in-situ experiments have been realized at a fixed temperature. Each temperature mentioned in this paper corresponds to a different experiment. The annealing temperatures are reached after an increase of 50°C/minute and then kept constant. The Raman spectra have been collected with acquisition times ranged from three to ten minutes, depending on the speed of the relaxation. The beginning of the isotherm, t=0, was taken at the time at which the temperature of equilibrium was reached. For very short relaxation times, this definition of the starting time could introduce an underestimation of the time. This point will be discussed later in the discussion section.

### C. In-situ density measurements

To allow in situ measurements of the density, a numerical program was developed to calculate sample surfaces. Optical pictures were taken just before the Raman acquisition and treated to extract the surfaces by numerical integration. Since the expansion is isotropic, it is possible to link the surface evolution to the volume, in order to deduce the densification ratio. Samples were platelet like shaped to minimize errors due to variations in the normal direction. Using the isotropic nature of the expansion, we followed the simple relation $V = S^{3/2}$ for the volume calculation. Each image outline was defined by the eye. To evaluate the error introduced by this method for each image, the analysis was realized several time, the overall standard deviation was determined and taken as error bar.

The flat and large Belt samples are particularly suited for this method. The errors obtained here less than 0.02 for densities are of the same order of magnitude than for previous reported technics, 0.01 for the lithographic mask method[23] or 0.01 for thickness measurement by integration of the transmittance.[8] However the samples densified in diamond anvil cell, around 50 microns size, are too small and their density could not be determined in a reliable way.

## III. RESULTS

### A. Raman spectra before and after annealing

The Fig. 1 shows the Raman spectra of recovered samples from DAC and Belt compression with similar densification ratios, and the Raman spectrum of the non-densified silica glass. The densified samples spectra show a narrowing and a shift of their main band at 450 cm$^{-1}$ toward the high frequencies compared to pristine glass. The main and the D1 bands merge rapidly with the increase of the densification rate. The main band shift to higher frequencies is more pronounced in the case of the DAC compression compared to the hot compression for similar densities whereas the main band low frequency half width at half maximum $\Delta L_{1/2}$, defined in Fig. 1(b), is independent of the (P,T) path densification. Moreover, the D2 band intensity centered at about 600 $cm^{-1}$, which corresponds to the breathing mode of three membered rings,[24,25,26] depends strongly on the compression temperature, as it is clearly shown on Fig. 1(a) in agreement with our previous work.[18] For the Belt samples obtained from hot compression, the D2 intensity decreases with densification, whereas an intensity enhancement has been observed after a cold compression in DAC.

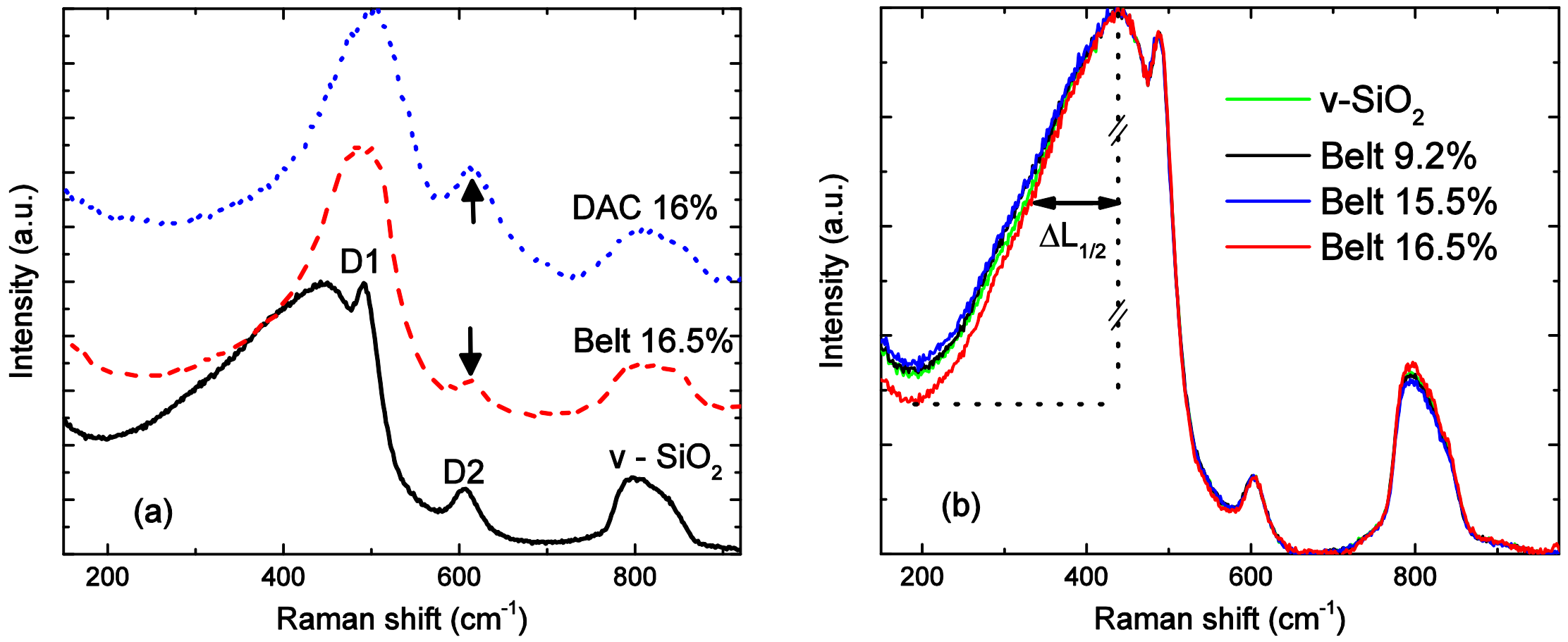

FIG. 1. (Color online) Raman spectra of non-densified silica glass and samples recovered from Belt and DAC compressions (a). Raman spectra of the Belt samples and the non-densified silica glass after an annealing at 1100°C for 2 hours (b).

In order to characterize the relaxations, a relevant structural parameter has been chosen. We used the mass center including the main band, the D1 and D2 bands, named σ and defined as[19]

$$\int_{200}^{700} f(x)dx = 2\int_{200}^{\sigma} f(x)dx \qquad (1)$$

It gives a good representation of the glass structure and shows a strong robustness to the presence of noise in the spectra due to its integration nature. This allows us to quantify the relaxation during an isotherm at temperature T as a function of the time t, by introduce transformation degree *y*, defined as

$$y(T,t) = \frac{\sigma(T,t=0)-\sigma(T,t)}{\sigma(T,t=0)-\sigma(T,t=\infty)} \qquad (2)$$

This expression can be simplified by applying the following considerations. A study in temperature of the pristine glass showed that its Raman spectra evolves in such a way from room temperature to 1000°C that its $\sigma$ value is independent of temperature. This independence of sigma with temperature is surprising. Usually due to anharmonicity the vibrational frequencies are expected to decrease with temperature. The position of the main band however, as observed by McMillan,[27] Le Parc[28] and ourselves, increases with temperature. The behaviour of the main band maximum is then abnormal. If the evolution of sigma is independent from temperature, that could probably be explained as the cross effect between abnormal evolution of the main band maximum and the overall normal evolution of the other vibrations.Indeed both abnormal and normal evolutions are present in the integration interval used to define sigma (see eq. 2).

Similarly, if we consider that no relaxation takes place during the fast heating of the densified glass up to T then $\sigma(T, t = 0)$ is the same than $\sigma(298\ K, t = 0)$.

In Fig. 1.(b) are presented the Raman spectra of the three Belt densified samples after annealing at 1100°C during 2 hours as well as the non-densified initial Suprasil 300 glass. One can see that all spectra are identical. The final relaxed structure is the same for the three densified states and for the non-densified silica. After thermal treatment all the full relaxed densified Belt samples present Raman spectra strictly similar to their initial glass. The same thermal treatment has been done both for the Belt and DAC samples. For all the samples, after a full thermal relaxation, the final values of the main band position and σ parameter are 436±1 $cm^{-1}$ and 412±1 $cm^{-1}$ respectively and the relative intensities of the D2 and D1 bands with respect to the main band are the same. The comparison of structural data and the spectrum of the final state shows that the densification (HP-HT) -relaxation cycle is fully recoverable, as it has been demonstrated for the polarizability,[29] the refractive index[29,13] and the density.[2] Then $\sigma(T, t = \infty)$ is the same than $\sigma(298\ K, pristine)$ and Eq. (2) can be rewritten as followed

$$y(T,t) = \frac{\sigma(298\ K,t=0) - \sigma(T,t)}{\sigma(298\ K,t=0) - \sigma(298\ K,pristine)} \quad (3)$$

This new expression allows to deduce the $y$ value. Nevertheless, the first Raman spectrum at t=0 or for the full relaxed state at t=∞ could not be acquired due to too fast or too slow transformation.

The only spectrum difference lies in the low frequency HWHM of the main band. The half width of the main band, denoted $\Delta L_{1/2}$ is related to the distribution of the Si-O-Si angle (see Fig. 1(b)). However after applying a baseline correction and integrating the all regions, these small variations do not affect the value of σ.

### B. Density evolution

It has been shown previously from ex situ data[18] that the density follows the main band width $\Delta L_{1/2}$. To verify this hypothesis during our study on structural transformations, we plotted the evolution $\Delta L_{1/2}$ versus density. As specified in the experimental part, surface calculations of flat samples were performed and we were able to obtain in situ densities. The temperature was increased step by step up to 1000°C to cover the full transformation range, and the Raman spectra were collected at each step. The data of Martinet[18] have been added. They correspond to identical silica (Suprasil 300), densified using the same methods, with different P-T parameters. From Fig. 2, density appears to follow the $\Delta L_{1/2}$ for the three different initial densities and form a master curve. However, it can be noticed that this master curve presents two linear behaviors below and above a density of 2.4. This slope change suggests that the structural evolution is not directly correlated to the density (see discussion). Nevertheless, for each sample, the density and the degree of transformation y have a monotonous relation. Even if $\Delta L_{1/2}$ could have been a better estimate of the density it gives only a partial estimate of the structure. Indeed, $\Delta L_{1/2}$ is related to the large ring Si-O-Si angle distribution whereas the degree of transformation is related to all the Si-O-Si motions. By integration on a larger set of value we end up with a parameter less specific but evolving more smoothly and presenting no abrupt change of slope.

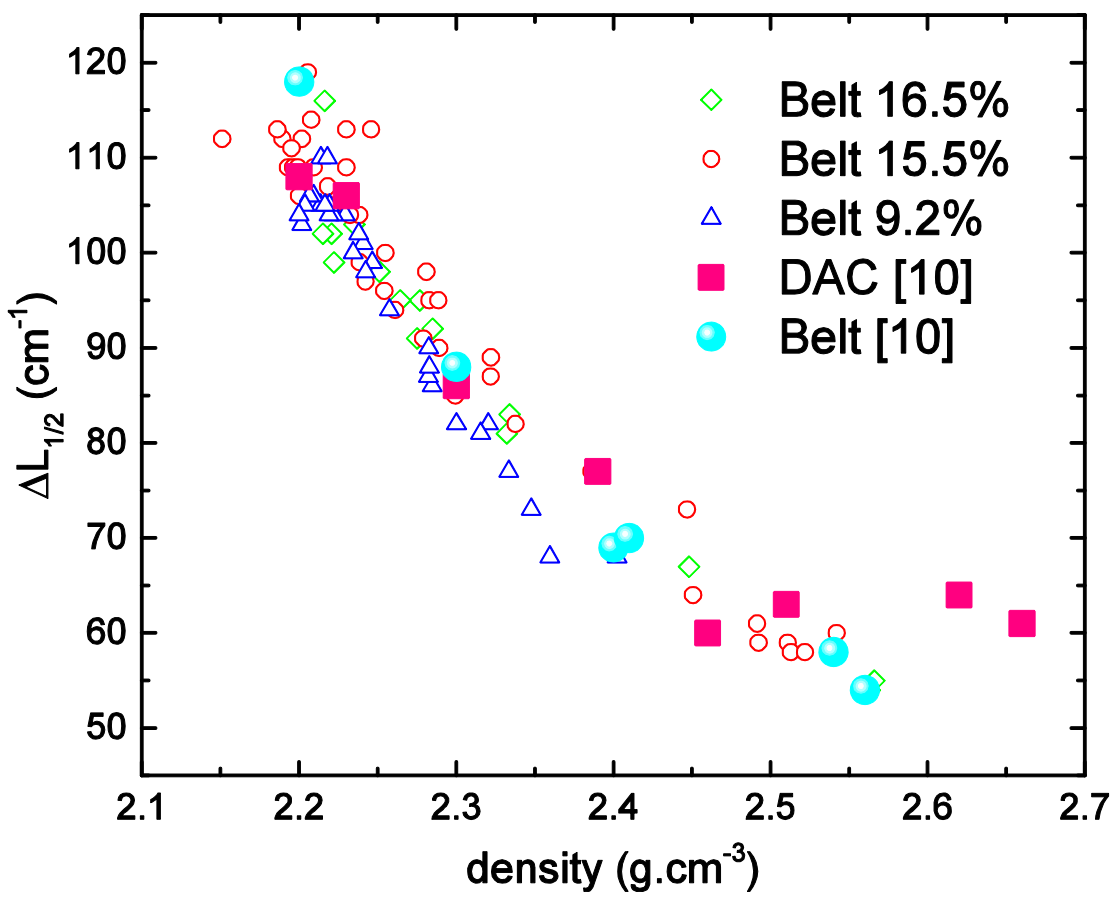

FIG. 2. (Color online) $\Delta L_{1/2}$ against density. The Ex-situ data added from Ref. 18 correspond to identical glass densified with the same methods at different P-T. Errors on $\Delta L_{1/2}$, and density are respectively 6 $cm^{-1}$ and less than 0.02 $g.cm^{-3}$.

## C. Structural evolution

For each densified sample, we performed a comparison of the Raman spectra with similar transformation rates, obtained after different annealing isothermal temperatures. Fig. 3(b) shows the evolution of the degree of transformation during different temperature annealing for the Belt 9.2% sample. Fig. 3(a) shows Raman spectra after an annealing of the Belt 9.2% sample at three different temperatures (700°C, 820°C, 850°C) when *y* is equal to 0.62. Once again, the indiscernibility of the three spectra allows us to claim that the glassy structures are identical for the same y value within the same compression (P,T) process. More than thirty-five other identical comparisons over a wide range of temperatures (500°C-900°C) and transformation rates (27%-100%) have been made on Belt and DAC samples, and the same result has been obtained. This means that for the same initial silica sample, which has been densified from high pressure and at a given temperature, whatever the annealing temperature, the structural evolution of this glass sample is the same for a given transformation rate, y. Then, this parameter can be used with confidence as a kinetic parameter.

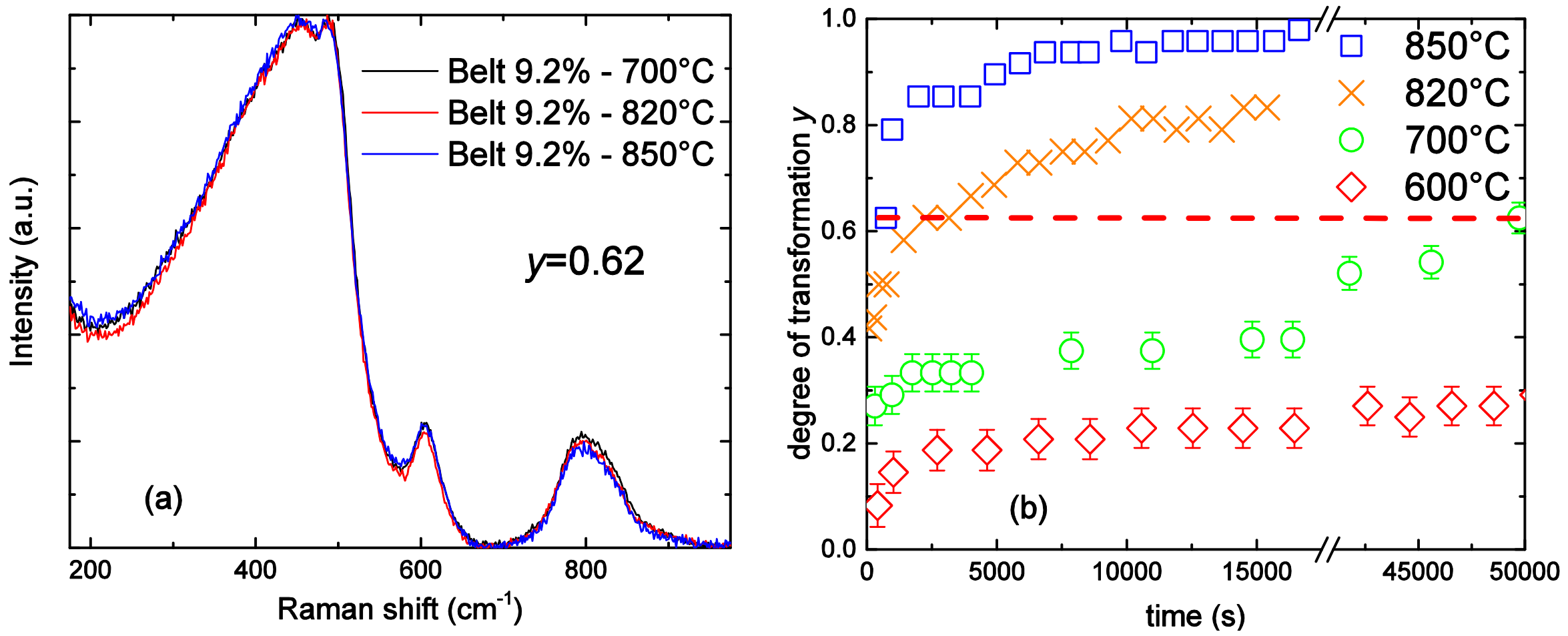

FIG. 3. (Color online) Raman spectra of three Belt 9.2% samples at different annealing temperature, at $y$=0.62 (a). Evolution of the degree of transformation for different annealing temperatures for the Belt 9.2% sample (b). The dashed line corresponds to $y$=0.62, the points on the line correspond to the data given in (a). The annealing temperatures are reported in the legend. Uncertainties not shown are contained within the size of the points.

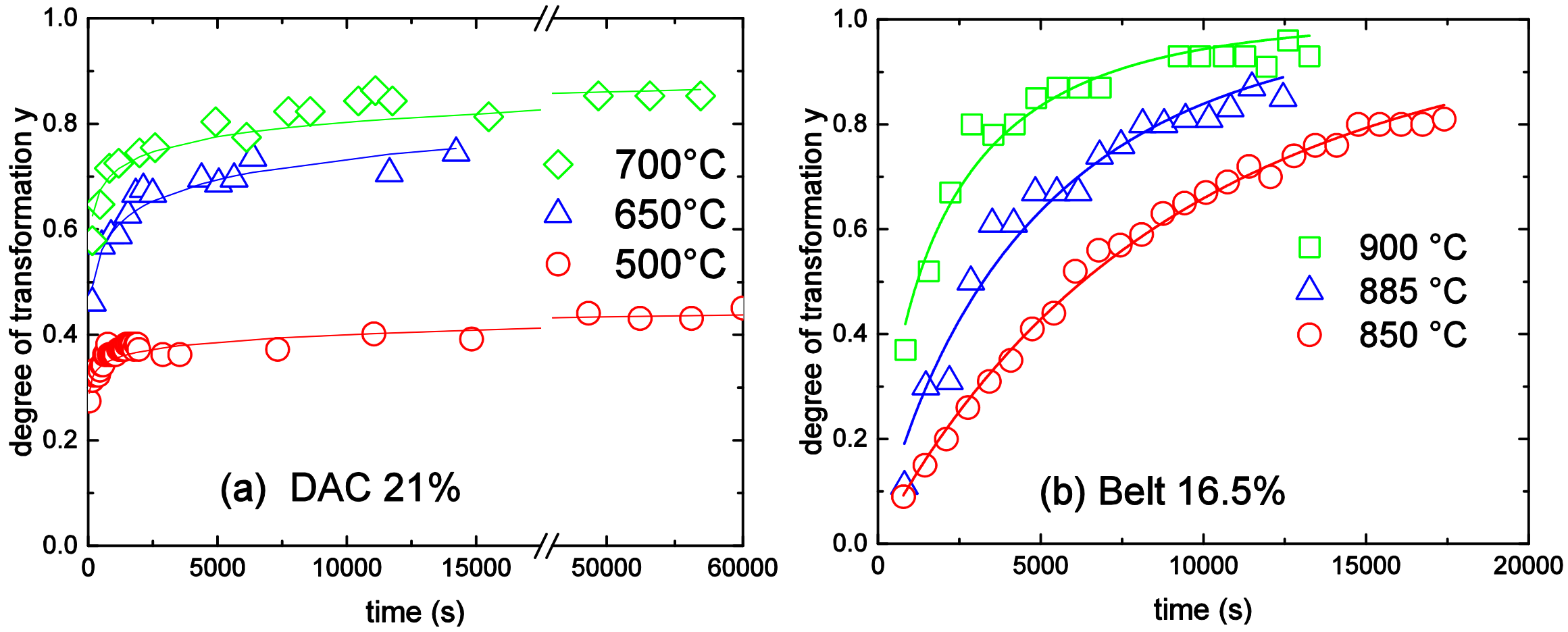

FIG. 4. (Color online) Evolution of the degree of transformation for different annealing temperatures for the highly densified samples: DAC 21% (a) and Belt 16.5% (b). The lines correspond to a stretched exponential fit. The annealing temperatures are inserted in the legend. Uncertainties are contained within the size of the points.

Different isothermal annealing were performed for all the densified samples (except for the DAC 16% sample). In Fig. 4 are presented the evolutions of the degree of transformation y extracted from the Raman spectra collected in-situ during the isothermal process for the Belt 16.5% and DAC 21% samples. For all the samples, the application of high temperature (500 to 900°C) allows to reach quickly a less dense silica glass, partially relaxed. After about $10^3$s for the DAC 21% sample and in a smaller extent after about $10^4$s for the Belt 16.5% sample, an apparent saturation on the curve appears in Fig 4. The presence of these apparent saturations means that the structure evolves so slowly that no further significant evolution can be observed in a reasonable time at the given temperature. At low annealing temperatures, i.e. 500°C or 600°C, compared to the $T_g$ of silica glass which is 1200°C, the structural evolution was very slow, and long-time measurements (about 15 hours) were realized, in order to extract a significant evolution.

We can also follow structural data, for example: $\Delta L_{1/2}$, band areas or band positions as a function of the degree of transformation. A special attention was given to the D2 band area evolution as a function of y. The D2 band area was normalized by the area under the Raman signal integrated between 200 $cm^{-1}$ and 900 $cm^{-1}$. For a same densified sample the data from the different isotherm can be plotted together (see low part of Fig. 5). Doing so, the obtained data for the different isotherms join together in a unique curve. The existence of these unique curves demonstrates that the structural relaxation path is not a function of the applied temperature and is then specific for each sample. For each sample, these D2 area curves as a function of y are reported on the upper part of Fig. 5(a) and Fig. 5(b). This allows us to compare the structural evolution of all our samples through the entire transformation.

We found out that the D2 line area undergoes a surprising non-monotonic behavior. Indeed, the D2 area increases at the beginning of the relaxation process and a maximum has been observed between $y = 0.1$ and 0.3 for all densified samples. This first increase was totally unexpected. At higher y the D2 line area decreases as a function of $y$ to reach back the value of the pristine glass as expected.

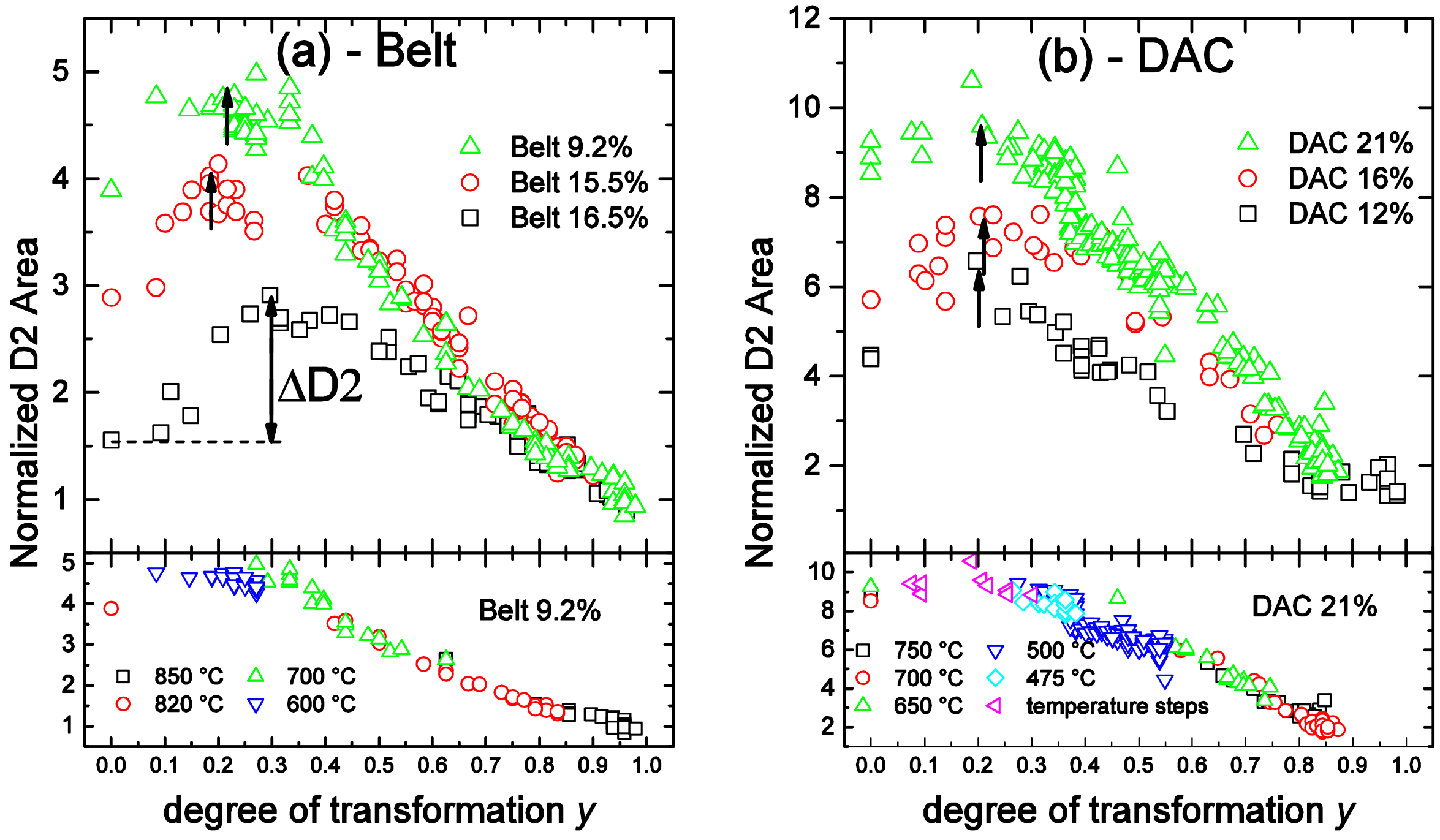

FIG. 5. (Color online) Evolution of the D2 band area against the degree of transformation for the Belt (a) and DAC (b) samples. The arrows mark the position of each maximum. The area is normalized by the total area under the Raman spectrum between 200 $cm^{-1}$ and 900 $cm^{-1}$. The curves presented on the upper part of the figure are obtained by superposing for each sample all the isotherms as exampled on the lower part.

Unfortunately, we were not able to follow the Raman active D1 band, relative to the four membered rings, since it was shown that only the pump probe technic is able to discriminate properly this band from the main band.[5,30]

The DAC 16% sample has been relaxed non isothermally. Indeed, the temperature was increased step by step up to 700°C and a spectrum was recorded at each step without waiting for a complete relaxation. Their D2 band areas are reported in Fig 5(b) with the data of the samples DAC 21% and DAC 12% obtained on isotherms. It can be seen that these data fit consistently between the two other series. This fact confirms that the annealing temperature does not change the structure at a given y.

## IV. DISCUSSION

Both methods of densification realised here are useful to understand more clearly the effect of pressure and temperature applied in the compression decompression cycle on the structural modifications. Because all samples were compressed at room temperature, the temperature relaxations on the DAC samples will highlight the effect of pressure on densification. In the other hand, the relaxation of the samples densified in the Belt press will give indications on the effects due to the applied temperature during the compression.

The variation of the compression parameters (P,T) allows the preparation of a wide variety of structure. A more exhaustive discussion about the structural differences between silica samples densified in Belt press and Diamond Anvil Cell can be found in previous papers.[18,29] Despite their structural differences our density study presented in Fig. 2(a) showed that the half width at low frequency of the main band, $\Delta L_{1/2}$, acquired during the temperature relaxation for the Belt samples is similar to that for previous DAC experiments[18] at equal densities. This new calibration curve between Raman spectra and density could provide a useful tool to estimate the density of silica samples, regardless of the temperature and pressure history.

From the calibration curve presented in Fig. 2 it can be seen that the decrease of the main band $\Delta L_{1/2}$ versus density presents two distinct regimes, a quick linear variation from 2.2 to 2.4 and a slower evolution from 2.4 to 2.7. The main band is related to the Si-O-Si bending mode of the group. The width of the main band decreases when the angle distribution decreases as a function of densification. The main band is asymmetric indicating that large intertetrahedral angles which correspond to the lower frequency contribution of the main band have a larger distribution than the small intertetrahedral angles. The value $\Delta L_{1/2}$ measures then more specifically the extent of the large angles contribution. In the first regime, between 2.2 and 2.4, the large angles seem to contribute strongly to the density evolution. In the second regime, between 2.4 and 2.7, large angles evolve only slightly with density and some other structural mechanisms take place as modification of ring sizes. The small ring populations can be observed on the high frequency part of the main band.[18,31] It is in this higher frequency region especially around the so call D2 band that the biggest differences were observed between low and high temperature densification. The D2 band which is assigned to three membered rings will be further discussed later. At the opposite the coefficient sigma, which is obtained by integration over the main band, the D1 and D2 contributions, smoothes the two regimes. Thecoefficient sigma is then more appropriated to construct the degree of transformation $y$ defined in Eq. 2.

In Fig. 4 are presented the evolution of the degree of transformation for the two most densified samples, Belt 16.5% and DAC 21%. The temperatures are 200°C to 400°C lower in the relaxation for the DAC samples. Moreover, the low transformation rates ($y$<20%) cannot be observed for the DAC densified samples in isothermal experiments, even for temperature as low as 500°C. So it seems that the DAC densified samples are less stable than the Belt densified ones, since glasses relax faster or at lower temperatures. It was shown that the samples densified through hot compression are more homogeneous.[10,31] Huang et al. showed with molecular dynamic numerical simulations that the chemical rearrangement events that are responsible for the irreversible structural changes are thermally activated processes.[10] Thus the proportion of such bond swapping events increases when the applied temperature during the compression increases. This implies that densification through compression at high temperatures allows the system to release continuously the stress generated by the high pressures, leading to a more stable densified, lower stressed and more homogeneous samples.

In order to extract relaxation times from our data, we used a stretched exponential function, also called Kohlrausch–Williams–Watts (KWW) function. As a common feature in glass science, the stretched exponential function (Eq. 4) provides accurate fit and satisfactory computation of the parameters.

$$y(t) = 1 - e^{-\left({t}/{\tau}\right)^{\beta}} \quad (4)$$

Examples of the fit result are presented in Fig. 4. A very good agreement within our error bars were observed for all the relaxation, even the fastest one observed for DAC samples. The parameters and their error bars (obtained with Levenberg-Marquardt fitting algorithm using origin software (OriginLab, Northampton, MA)) are reported in Table II. β corresponds to the stretching exponent and $\tau$ is the relaxation time.

However, the parameter β, comprised between 0 and 1, has unclear origin and signification. Consequently, efforts have been made in order to provide a microscopic origin to the stretched exponential function. In particular, the model of diffusion of excitations into traps, developed by Grassberger and Proccacia[32] has received experimental support by Philipps[33,34] for a wide variety of relaxations phenomena. This model states that the stretched exponential has a topological origin, with relaxation phenomena governed from either short or long range interactions, leading to two specific values for the stretched exponent, respectively 3/5 and 3/7. The values of β plotted against the initial densification in Fig. 6 show a continuous distribution in the all interval β=0 to β=1, with no appearance of any magic value. Here we see clearly that this model is inadequate to explain our data.

In another point of view, β is usually interpreted as a consequence of the inhomogeneity of the glassy material. Indeed, the stretched exponential can be rewrite as follow[34]

$$y = 1 - \int \rho(u) e^{-t/u}\, du \qquad (5)$$

From Eqs. (4) and (5), the β exponent reflects the distribution *ρ(u)* of the characteristic relaxation times $u$, distribution that arises from the different local environments of the relaxation events. A β equals to one corresponds to the classic Debye relaxation, and the inhomogeneity of the transformation increases when the β values tends to zero. The values of the parameter β for all the isothermal annealing performed in our study, range from 0.06 to 0.92 and are listed in Table II. These values are plotted against the initial densification ratio in Fig. 6. The β values are really low, below 0.3 for the DAC densified samples. They tend to decrease with the maximum applied pressure during synthesis. This small decrease of beta suggests that the inhomogeneities increase with the applied pressure at room temperature for DAC compression. The Belt samples all densified at 5 GPa and at different temperatures, are then only sensitive to compression temperature. The temperature increase during the compression induces a strong β increase. Beta is even very close to one for a compression performed at 1020°C. Thus, as suggested by the β variation, the effect of temperature confirms the relative better homogeneity of the belt samples as discussed above. The beta coefficient is related to the homogeneity of the relaxation, which is directly correlated to the initial homogeneity of the densified samples. It can also be noticed that the samples named "Belt 9,2%" and "DAC 12%" have a similar β coefficient.

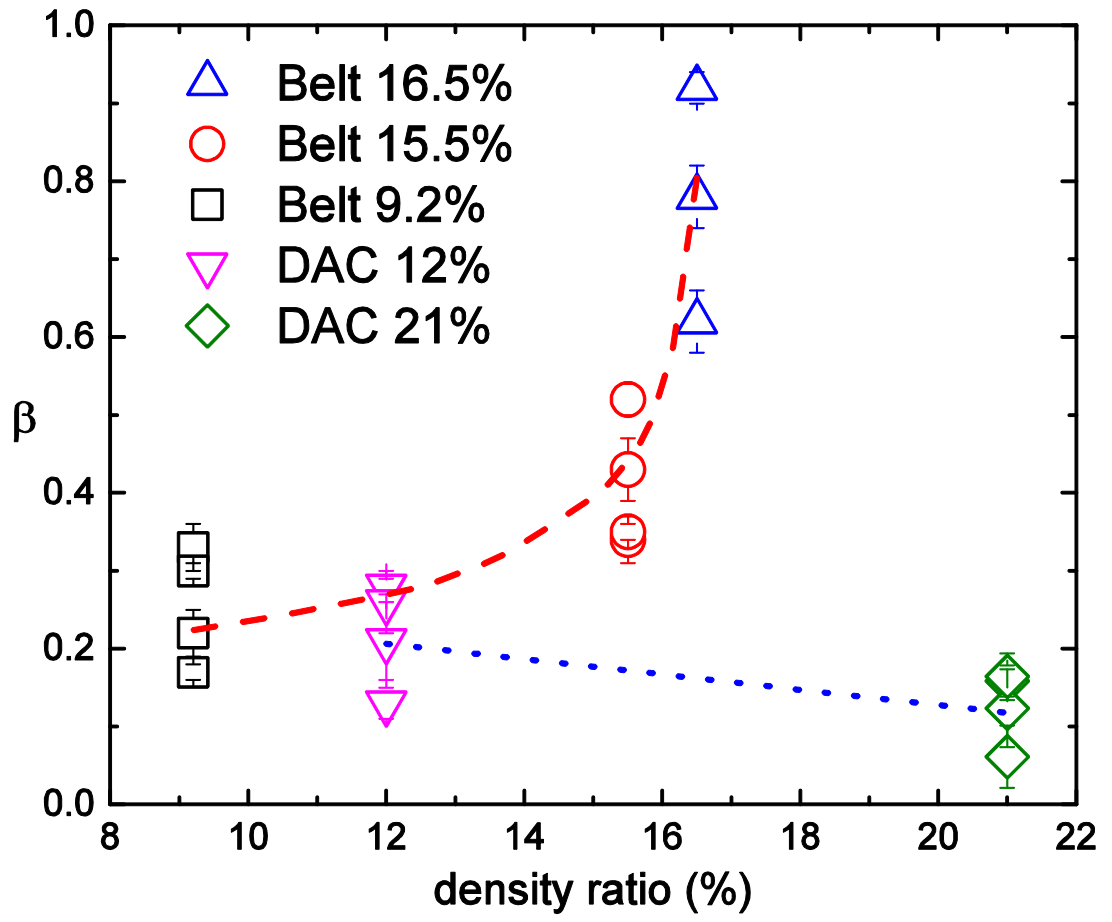

FIG. 6. (Color online) β values deduced from Eq. (3) for all the isothermal annealing performed in our study against the initial densification ratio. Dashed and dotted lines are guide for the eyes for Belt and DAC samples respectively.

It can be seen from Table II that the relaxation time $\tau$ can change by several orders of magnitude within temperature interval of less than 100°. That suggests an Arrhenius behaviour (Eq. 6).

$$\frac{1}{\tau} = Zexp\left(-{}^{E_a}/_{RT}\right) \quad (6)$$

At this point, one should take in consideration that the stretched exponential stands for a distribution in relaxation time. Thus the characteristic time to enter in the Arrhenius formula is the average relaxation time ⟨τ⟩ while the τ parameter given by the fit is the most probable relaxation time. The average relaxation time is defined as :[35]

$$\langle\tau\rangle = \frac{\tau}{\beta}\Gamma\left(\frac{1}{\beta}\right) \quad (7)$$

Here Γ is the gamma function. An immediate consequence of this definition is a very large increase for ⟨τ⟩ as β decreases, with the same trend for the errors. Indeed the errors on ⟨τ⟩ are larger than the nominal value for all the points from the relaxation of DAC samples. Thus the computation of the activation energies becomes non rigorous for these samples, and we focused on the Belt samples in this part. We plotted in Fig. 7 the structural relaxation times against the reciprocal temperature. For each sample the values are aligned and the relaxation time as a function of inverse of temperature is linear, confirming Arrhenius behaviour. The slopes of the three Belt samples are similar suggesting a unique relaxation phenomenon. Then all the experiments were fitted together. The activation energy obtained is 255±45 kJ/mol.

The activation energy for viscous flow at these temperatures[36] (below 1400°C) is 720 kJ/mol (7.5 eV). The great difference between this value and the one found in our study indicates that the structural changes responsible for the density relaxation are different than the ones associated with viscous flow. Thus the structural modifications implied here are fundamentally different from the classical viscous structural relaxation, i.e. the relaxation of a glass close to the glass transition temperature.

Few references can be found in the literature to compare other values of $E_a$ in the case of pure silica glass. Similar relaxations of densified silica samples have been carried, but mostly to explain the density versus refraction index behaviour,[29] with no kinetics considerations. Höfler and Seifert,[37] using density measurements, suggest $E_a$ = 290 kJ/mol (3 eV) for silica densified at 900°C and 2 GPa, i.e. for an initial densification ratio of 7% similar to our belt experiment. This value was also calculated using KWW function, but done using the time parameter obtained in the fit directly, so without reducing it as ⟨τ⟩.

Thus the $E_a$ value given by Höfler[37] need to be corrected, which gives 300 kJ/mol with unknown uncertainties. Considering similar error as ours, this value is in good agreement with our result.

Many authors use the Eq. 6 without considering $\langle\tau\rangle$.[37-40] This common practice is not a problem if the β values are almost constant and range from 0.5 to 1. In that case $\Gamma(1/\beta)/\beta$ is comprised between 1 and 2, and their corrected result remains within the errors bar.

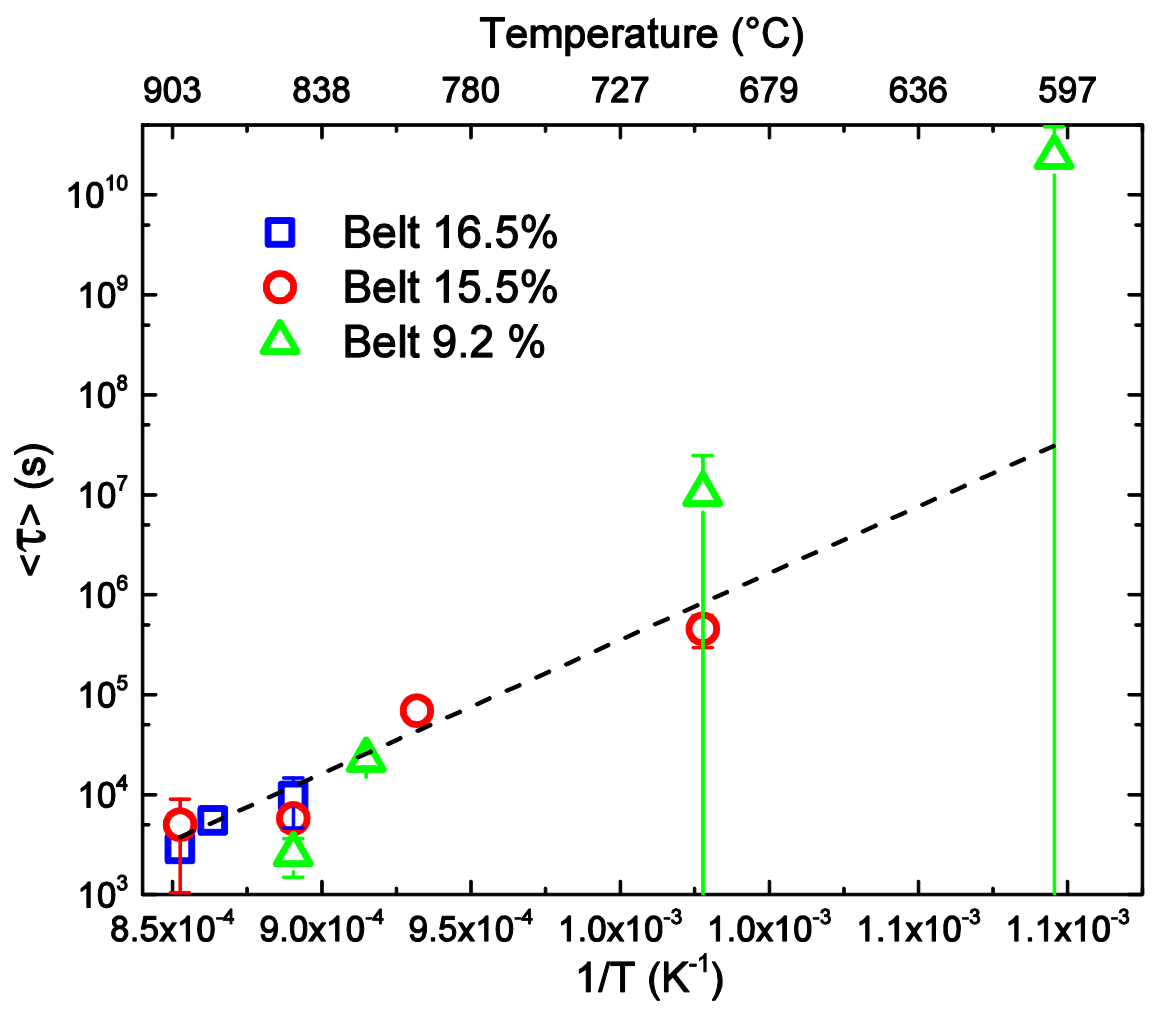

FIG. 7. (Color online) Structural relaxation times against the annealing temperature. The dashed line is the linear fit to the data excluding the last two points of the Belt 9.2% sample. Uncertainties not shown are contained within the size of the points.

For the reasons listed above, the calculation of the activation energy of the relaxation for the cold (DAC) densified samples is impossible. Nevertheless, qualitatively one can notice that for these samples the relaxations take place at lower temperatures compare to the hot (Belt) compressed samples. Even for the Belt samples, one can see from Table II that the minimum annealing temperature gets lower with the compression temperature. Furthermore we observed that after 10 minutes at 150°C a 10% relaxation was realized for the DAC 21% sample. In good agreement with this observation, Grimsditch[41] showed that silica densified at room temperature starts to relax for temperature as low as 100°C, suggesting a very low activation energy. We attribute this difference in kinetic between cold and hot densified samples to the fact that internal stresses help the relaxation. As stated out by Johari,[42] plastically deformed glasses retain energy as internal stresses that can be later released on heating. In a glass, these internal stresses are probably arising from the mismatch of the elastic modulus and/or density fluctuations. Thus they are directly related to the homogeneity of the glasses. As discussed earlier, the hot compression leads to more homogeneous glasses, with high β value, less internal stresses and then presenting slower kinetic.[18,31] Koziatek et al. computed the distribution of activation energies in silica glass after a cold compression at 25 GPa (corresponding to the DAC 21% samples) using molecular dynamics.[43] They found an asymmetric distribution of activation energies centered on 1.5 eV (145 kJ/mol, with a width in the high energy side of 1.5 eV). This is consistent with the fastest kinetic observed on our DAC samples. In this study, at least a minor part of the relaxation is released during the heating ramp. It is necessary to evaluate this part on the determination of the activation energies. Using back the kinetic coefficient fitted, it was possible to estimate the equivalent relaxation time reached during the heating ramp. For all the experiment a delay between 25 and 50 s was found. Thus, we assume that the errors arising from this unfollowed relaxation part are within the errors determined by the fit (see error bars reported in Table II).

Table II. Summary of the fitted values for the β parameters, the relaxation time τ and the activation energies $E_a$ for the different initial silica samples. The activation energies are not determined (n.d.) for the DAC samples due to too large errors (see text)

| Samples | Compression parameters | Annealing temperatures (°C) | β | τ(s) | Activation energies ($KJ.mol^{-1}$) |
|---|---|---|---|---|---|
| Belt 16.5% | 5 GPa, 1020°C | 900 | 0.62±0.04 | 1990±150 | 255±45 |
| | | 885 | 0.78±0.04 | 4770±160 | |
| | | 850 | 0.92±0.02 | 9300±100 | |
| Belt 15.5% | 5 GPa, 750°C | 900 | 0.34±0.03 | 900±100 | 255±45 |
| | | 850 | 0.52±0.02 | 3067±102 | |
| | | 800 | 0.35±0.01 | 13760±644 | |
| | | 700 | 0.43±0.04 | $(165±55).10^3$ | |
| Belt 9.2% | 5 GPa, 420°C | 850 | 0.33±0.03 | 410±80 | 255±45 |
| | | 820 | 0.3±0.01 | 2430±130 | |
| | | 700 | 0.22±0.03 | $(180±54).10^3$ | |
| | | 600 | 0.17±0.01 | $(41,6±27).10^6$ | |
| DAC 21% | 25 GPa, 25°C | 750 | 0.16±0.02 | 210±70 | n.d. |
| | | 700 | 0.12±0.05 | 175±60 | |
| | | 650 | 0.16±0.03 | 1730±100 | |
| | | 600 | 0.24±0.04 | 3770±120 | |
| | | 500 | 0.06±0.04 | $(9±4).10^6$ | |
| DAC 12% | 13.8 GPa, 25°C | 800 | 0.26±0.04 | 25±14 | n.d. |
| | | 700 | 0.28±0.01 | 3335±300 | |
| | | 600 | 0.21±0.05 | $(170±170).10^3$ | |
| | | 500 | 0.13±0.02 | $(500±290).10^3$ | |

The kinetic study only gives a very rough idea of the structural evolution during the relaxation. At this point the structural origin of the activation energy measured is still unclear. To provide a more complete picture of the structural evolution, the two different spectroscopic signatures: $\Delta L_{1/2}$ and the D2 area were followed.

We have been able to connect with the density while the D2 band arises from the collective breathing mode oxygen atoms within the 3 membered rings in the silica network. The connection between the D2 band and the 3 membered rings was made by Galeener in 1983.[44] Further efforts to quantitatively estimate the link between the D2 area and the proportion of 3 membered rings have been made on the 2000's, based on ab initio numerical simulations. Umari, Pasquarello et al.[24,25] concluded that the area under the D2 band in the Raman spectrum is directly proportional to the population of 3 membered rings. This point of view has been followed in several experimental studies, in particular those focused on the D2 (and D1) band.[26,30,45] Nevertheless, Rahmani et al. calculation indicated that all the breathing modes of the 3 fold rings do not participate to the D2 band,[46] even if most of them do. So the D2 band arises effectively from the collective breathing mode of 3 membered rings, but give a low estimate of its effective total content. The different conclusions might come from the fact that the box simulated in botharticles are to small: 72 atoms for Umari and Pasquarello, and three configurations of 26 $SiO_2$ units for Rahmani. Indeed, as shown by Rahmani, the participation of small membered rings to the defect band is not the same for all the configurations. Better statistics are needed to get unambiguous conclusions. The intensity of the D2 band is also determined by the Raman coupling-to-light coefficient. Umari et al. demonstrated that this coefficient is related to the cosine of the Si-O-Si angle. Since the D2 shifts only slightly under pressure or during the annealing, this coefficient remains the same, as verified by B. Hehlen.[26]

Thus, if one cannot know the absolute proportion of the 3 membered rings, an evolution of the D2 band area truly reflects an evolution in the proportion of such rings in the glass

Even if the D2 band area cannot be used to determine an absolute proportion of 3 memebered rings, the D2 band area evolution shows the trend of the 3 membered rings population.

As shown in Fig. 5, the evolution of the D2 area during the relaxation is non-monotonic, and exhibits a maximum **for all the studied samples**. This result shows a complex behaviour of small rings population during the transformation: a first process where the creation of such rings dominates, and a second process where the proportion of three membered rings decreases. This increase of 3 membered rings in the beginning of the relaxation demonstrates the existence of a transitory state, necessary to achieve the density expansion. This implies non trivial bond redistribution during the transformation, and a complex recovering of the pristine glass. Furthermore, the universality of this behaviour for all the samples studied tends to indicate that this step of the relaxation mechanisms is common to the different samples, regardless the compression parameters. To characterize further the role of the creation of the 3 membered rings during the relaxation process, the behaviour of the D2 band can be detailed. In Fig. 8 we plotted the D2 band area of the densified DAC and Belt samples and the difference between this last value and the D2 band area maximum value, ΔD2, as described in Fig. 5.

These data are represented as a function of the averaged β values from all the annealing temperatures for each initial densification. Doing so, the aim is to test if some links can be established between the initial inhomogeneities due to the presence of three membered rings (initial D2 band area), the transitory state (increase in D2 band area Δ) and the inhomogeneities associated to the relaxation time distribution (coefficient β).

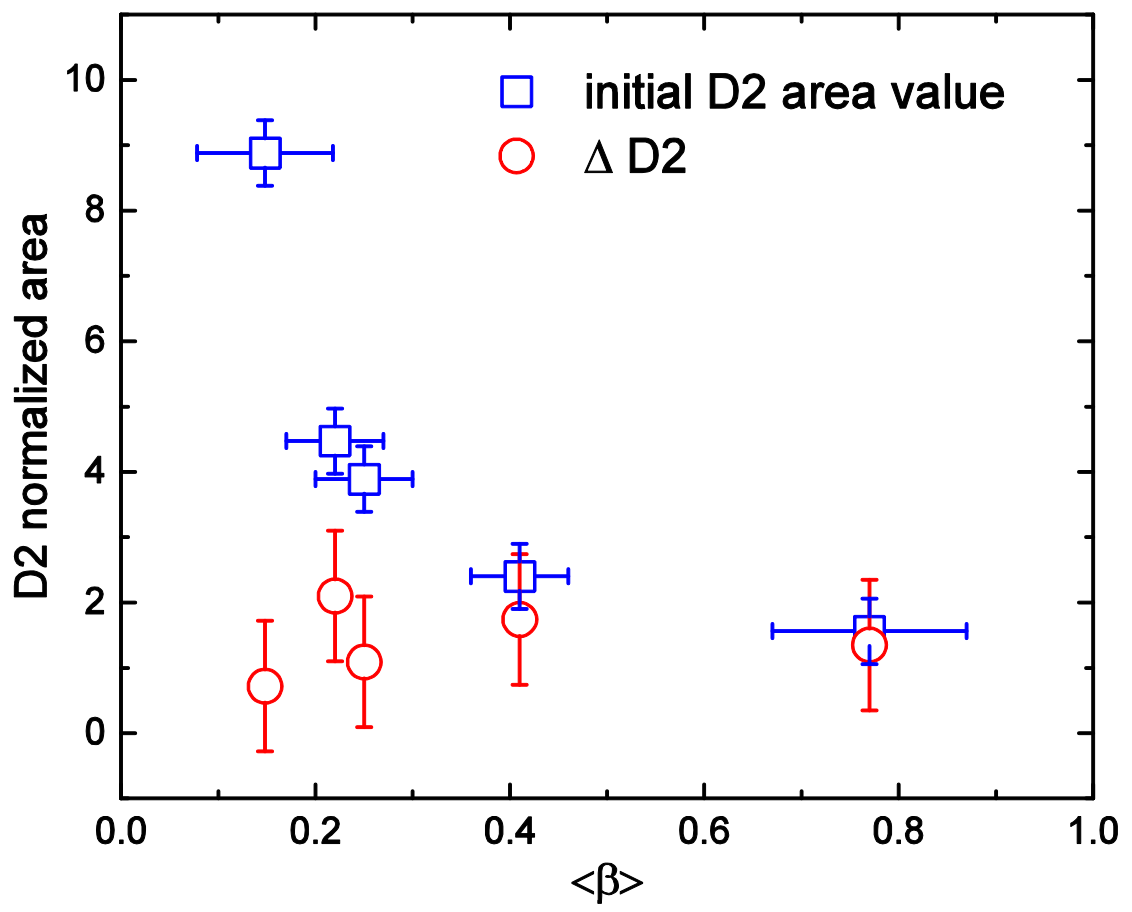

FIG. 8. (Color online) Initial D2 area values (blue squares) and gaps between initials and maximum D2 area, ΔD2 (red circles) against the averaged β. The averaged β are the mean values from all the relaxation performed on each initial densification e.g., Belt 16.5%.

From Fig. 8, the initial D2 band area follows the evolution of the ⟨β⟩ for all the samples. Indeed, the <β> goes from 0.15, large relaxation time, to 0.8, narrow relaxation time distribution when D2 area decreases. The inhomogoneities related to the time distribution are correlated to the importance of the three membered ring population in the densified samples. Thus, the same structural changes seem to characterize both evolutions. However, while the cold compression induces the creation of more 3 and 4 fold rings[18] related to the whole ring distribution in the densified samples, ΔD2 remains the same. Thus ΔD2 does not depend on the initial concentration of three membered rings. An extra population of 3 membered rings is always needed to initiate the structural relaxation process.This demonstrates that the initial population of 3 membered rings and the added one are two different populations. The initial population does not initiate the relaxation process.

The 3 membered rings are one of the most compact structure that can be found in the medium range order. Their increase should induce a densification however we observed a monotonous density decreases. Although counter intuitive, the creation of the 3 membered rings is not a consequence of the relaxation but it is actually a necessary step of “transitory states”. In other words, it could be understood as a volume of activation. The created 3 membered rings increase the density fluctuations, leading to higher contrast of interstitial volumes where densification relaxation occurs trough an opening of the Si-O-Si angles and changes in the dihedral angle.

More than only related to $SiO_2$, the existence of this transitory state seems to confirm the existence of an energy barrier between the densified and pristine glasses. In a landscape theory approach that means that densified and pristine glasses are inside separate mega basins. Following Denis Machon[47] the existence of separate megabasin could be a confirmation of the polyamorphism theory of glasses.[48,49]

It seems also that a general vision of the difference between cold and hot densification can be here deduced. In very good agreement with the conclusion of Johari,[42] the existence of strong internal stress is favoured at low temperature.

## V. CONCLUSION

Densification relaxation abled us to get many glasses with different densities. It was found that a good correlation could be found between the low frequency half width at half maximum and the density. This correlation could be a promising calibration between Raman signatures and density since it remains valid whatever the temperature and pressure history. The comparison between the two paths of densification allowed us separating the effect of pressure and temperature. The main difference was found in the stretching coefficient $\beta$ which is related with sample homogeneity. It can be then interpreted that the use of high temperature during densification introduced homogeneity in the glass. More the initial glass is homogeneous more it will take time to relax at a given temperature. All the kinetic terms were determined satisfactory for the first time for the samples densified at high temperature. The dominant role of internal stresses contribution in the relaxation has been put in evidence for the cold densified samples. The observation of a transitory state during relaxations was unexpected and seems to indicate that very local structural rearrangements rule the overall process. Even if the density relaxation is a continuous phenomenon local density heterogeneities are needed to allow an overall relaxation. A direct observation of these inhomogeneities could be done by tracking density fluctuations using small angle X ray scattering or Landau Placzeck ratio. Such measurements will confirm our observed transitory state.

## ACKNOWLEDGMENTS

The authors are thankful to the vibrational spectroscopies Platform at University Lyon 1 France (CECOMO) facility for micro-Raman spectroscopy experiments. The authors wish also to thank the french-german Procope program (project 35422YC) for the exchanges between France and Germany.